# Unconventional ratiometric-enhanced optical sensing of oxygen by mixed-phase TiO$_2$

S. Lettieri[1*], D.K. Pallotti[1,2], F. Gesuele[2] and P. Maddalena[2]

[1]*Institute of Applied Sciences and Intelligent Systems "E. Caianiello", CNR-ISASI, Via Campi Flegrei 34, I-80078 Pozzuoli, Napoli, Italy.*

[2]*Physics Department "E. Pancini", University of Napoli "Federico II", Via Cintia, I-80126 Napoli, Italy.*

**Abstract**. We show that mixed-phase titanium dioxide (TiO$_2$) can be effectively employed as an unconventional, inorganic, dual-emitting and ratiometric optical sensor of O$_2$. Simultaneous availability of rutile and anatase TiO$_2$ PL and their peculiar "anti-correlated" PL responses to O$_2$ allow using their ratio as measurement parameter associated to O$_2$ concentration, leading to an experimental responsivity being by construction larger than the one obtainable for single-phase PL detection. A proof of this concept in given, showing a two-fold enhancement of the optical responsivity provided by the ratiometric approach. Besides the peculiar ratiometric-enhanced responsivity, other characteristics of mixed phase TiO$_2$ can be envisaged as favorable for O$_2$ optical probing, namely: a) low production costs, b) absence of heterogeneous components, c) self-supporting properties. These characteristics encourage experimenting its use for applications requiring high indicator quantities at competitive price, possibly also tackling the need to develop supporting matrixes that carry the luminescent probes and avoiding issues related to the use of different components for ratiometric sensing.

Determining the content of molecular oxygen (O$_2$) in gas mixtures and/or aqueous environments is important in many different fields. For example, monitoring the O$_2$ concentration in the exhaust stream of an internal combustion engine provides information on the air-fuel ratio of the combustion process, whose control is relevant for reducing harmful emission[1–3]. Knowing the concentration of O$_2$ dissolved in water (also "dissolved oxygen", DO) is also crucial in many applications. DO level is intimately related to the state of health of marine habitats, to the concentration of organic pollutants in water[4] and is of great importance for biological studies. In fact, DO content in cell culture media affects cell growth rate, metabolism and protein synthesis, exerting a decisive influence on the cellular functions and behaviors. To mention a few examples, hypoxia/re-oxygenation induce stem cell like phenotype in prostate cancer cells[5] and enhance the proliferation, invasiveness and metastatic potential of tumor cells[6,7], while hyperoxia can cause cell death via the formation of reactive oxygen species[8,9].

---

* Corresponding author. E-mail: s.lettieri@isasi.cnr.it

Molecular oxygen in gas phase is in most cases detected by electrochemical or chemoresistive gas sensors based on metal oxide (MOX) semiconductors, such as zirconium dioxide ($ZrO_2$) or titanium dioxide ($TiO_2$). For a long time, Clark-type electrodes[10] have been considered the conventional devices for DO measurements, thanks to their robustness and reliability. However, technologies based on optical measurement of DO have received increasing attention in last three decades, as they do not suffer from electrical interference, do not consume oxygen and can be rather inexpensive and easy to miniaturize[11].

The most diffused and precise $O_2$ optical sensors are based on changes in photoluminescence (PL) lifetime of a sensitive material. This approach is not affected by the fluctuations in the excitation light intensity, but also requires more complex interrogation systems (e.g. optical phase detection). A simpler and more direct optochemical sensing method is based on the quenching of PL intensity of a sensitive material caused by the presence of $O_2$ according to the Stern-Volmer relationship[12].

While organic dyes are usually employed as $O_2$-sensitive photoluminescent materials, it is worth noting that other materials, such as porous silicon and different metal oxides, also exhibit environment-dependent PL emission and are indeed under scrutiny for possible application in optochemical sensing.[13–23]

Unfortunately, approaches based on intensity measurement of the PL emission are also prone to errors. In particular, fluctuations in the excitation light intensity can hinder the understanding of the experimental data when real-time monitoring is performed and small integration times in the PL measurement have to be used. Therefore, efforts have been focused in the last years to establish methods using an internal PL reference. These methods are also known as "ratiometric", as they are based on the acquisition of the ratio between the PL intensity of an $O_2$-sensitive species ("indicator") and the PL intensity of an $O_2$-insensitive species. As far as the two PL intensities are both linearly proportional to the excitation light intensity, the mentioned fluctuations are canceled out in the ratio, thus improving the experimental precision[24].

Ratiometric optical sensors are often realized by combining two different elements (typically, two dyes) as indicator and reference. This however involves issues related to unequal stability, reliability and photobleaching of the two different components[25]. To tackle these problems, it is therefore useful to purse and investigate PL-based sensing approaches based on some dual-luminescence material[26]. In this work we show that mixed-phase titanium dioxide ($TiO_2$) can be effectively employed as an unconventional, inorganic, dual-emitting and ratiometric optical sensor of $O_2$ exhibiting a peculiar "*ratiometric-enhanced responsivity*". We first illustrate the basic idea, giving next an experimental proof-of-concept.

Our starting point is observing that $TiO_2$ exhibits a *crystalline phase-dependent* PL activity, *i.e.* that the PL spectra associated to the anatase and rutile $TiO_2$ phases are well separated[27] and can be easily and unambiguously singled out experimentally. In particular, the PL of rutile-phase $TiO_2$ extends in the near infrared (NIR) spectral region (~780-920 nm wavelength interval, PL maximum at about 840 nm), while that of anatase-phase $TiO_2$ lies instead in the visible (VIS) range and is quite broad, extending from about 450 nm to 700 nm with a relative maximum around 500-520 nm wavelength. Both VIS-PL and NIR-PL intensities are affected by interaction of $TiO_2$ with $O_2$. Most importantly, *they change in opposite directions during interaction with oxygen*[28], i.e. the VIS-PL intensity decreases when (anatase) $TiO_2$ is exposed to $O_2$ (PL quenching), while the NIR-PL intensity of (rutile) $TiO_2$ increases instead (PL enhancement) when it interacts with $O_2$. Such a crystalline phase-dependent and "anti-correlated" behavior of $TiO_2$ PL can be exploited to increment the sensitivity toward $O_2$ through the *simultaneous detection of both* (VIS and NIR) *PL intensities* and by *use of their ratio* as optical transduction signal.

To illustrate the point in a quantitative fashion, we find it useful to define here a suitable parameter quantifying the optical responsivity and discuss our idea in terms of responsivity increment. In general, the PL response towards a given analyte can be quantified through the relative modulation of the PL intensity $R = |\varphi - \varphi_0|/\varphi_0$, where $\varphi_0$ and $\varphi$ are the steady-state PL intensity in absence and in presence of the analyte (here $O_2$), respectively. However, a so-defined responsivity is always less than one (or, equivalently, less than 100%) in PL quenching ($\varphi < \varphi_0 \Rightarrow (\varphi_0 - \varphi)/\varphi_0 < 1$), while it can be larger than one (larger than 100%) for PL enhancement. Thus, it does not allow a simple comparison between these two cases. For the sake of comparisons, it is instead preferable defining a dimensionless responsivity whose values are always limited in the interval (0,1) (equivalently, in the range 0% - 100%) for both the quenching and enhancement cases. This can be obtained by using the quantity

$$R = |\varphi - \varphi_0|/\max(\varphi, \varphi_0) \qquad (1)$$

for assessing the PL responsivity, where max ($\varphi, \varphi_0$) represents the largest quantity between $\varphi$ and $\varphi_0$. Using $\varphi_V$ and $\varphi_N$ to express the total VIS-PL and NIR-PL intensities (respectively), the responsivities $R_Q$ and $R_E$ associated to PL quenching and enhancement (respectively) and defined according to Equation (1) are therefore:

$$R_E = \left(\varphi_N - \varphi_N^0\right)/\varphi_N \qquad (2)$$

$$R_Q = \left(\varphi_V^0 - \varphi_V\right)/\varphi_V^0 \qquad (3)$$

It is easily seen that the above responsivities are, by definition, always less than one ($R = 1$ corresponds to the purely ideal cases of $\varphi_V = 0$ for PL quenching and $\varphi_N = +\infty$ for PL enhancement).

As mentioned previously, "ratiometric-enhanced" responses now arise by taking the PL intensity ratios as the transduction signal. In terms of responsivities (indicated here by using a long accent, i.e. $\bar{R}_E, \bar{R}_Q$) this means substituting ($\varphi_N/\varphi_V$) in place of $\varphi_N$ in Equation(2) and substituting ($\varphi_V/\varphi_N$) in place of $\varphi_V$ in Equation (3). We thus obtain:

$$\bar{R}_E = \frac{\left(\varphi_N/\varphi_V\right) - \left(\varphi_N^0/\varphi_V^0\right)}{\varphi_N/\varphi_V} \qquad (4)$$

$$\bar{R}_Q = \frac{\left(\varphi_V^0/\varphi_N^0\right) - \left(\varphi_V/\varphi_N\right)}{\varphi_V^0/\varphi_N^0} \qquad (5)$$

By developing the fractions in Equations (4) and (5), we see that the two quantities actually coincide, leading to a unique "ratiometric" responsivity $\bar{R} = \bar{R}_E = \bar{R}_Q$ given by:

$$\bar{R} = 1 - \left(\frac{\varphi_N^0}{\varphi_N}\right)\left(\frac{\varphi_V}{\varphi_V^0}\right) \qquad (6)$$

It is easily seen that such a modified PL ratio-based responsivity is by construction larger than the ones ($R_Q, R_E$) obtainable using single-phase $TiO_2$ (i.e. when only one of the two possible PL bands is available). In fact, using Equations (2), (3) and (6) the following equalities can be obtained:

$$\begin{aligned}\Delta R_E &= \bar{R} - R_E = R_Q\left(1 - R_E\right) \\ \Delta R_Q &= \bar{R} - R_Q = R_E\left(1 - R_Q\right)\end{aligned} \qquad (7)$$

Recalling that the responsivities $R_Q$ and $R_E$ are (by definition) less than unity, we see that the $\Delta R$ quantities are *always positive*, i.e. $\bar{R} > R_E$ and $\bar{R} > R_Q$. We thus conclude that the simultaneous availability of anatase and rutile $TiO_2$ PL and the detection of their ratio as transduction signal allows reaching an experimental responsivity that is always larger than the one obtainable for single-phase PL detection. This illustrates the concept to which we referred to as "ratiometric-enhanced responsivity".

In order to give a proof of this concept, we prepared and tested a mixed phase $TiO_2$ ("MPT" hereafter) sample and two reference single-phase (rutile and anatase) samples. They were prepared starting from commercial anatase nanopowders (Sigma Aldrich 232033) and Aeroxide® P25 nanopowders (Sigma Aldrich 718467). The anatase powder was annealed in ambient air at 300°C to remove eventual organic contaminants, while P25 powder was annealed at 900°C in ambient air for 1 hour, guaranteeing its full crystallization in the rutile phase. The powders were cold pressed by a hydraulic press to obtain pellets of 13 mm diameter and about 0.5 mm thickness. MPT pellets were prepared by mixing two phases powders (rutile/anatase proportion 2:1 in mass) while reference anatase and rutile samples were obtained using the individual single-phase powders. All the pellets were analysed through PL and Raman spectroscopy.

Raman characterization was performed through a home-built optical setup using a He-Ne laser (633 nm wavelength, 5 mW power) as excitation source and a microscope objective (10x magnification, 0.25 numerical aperture) coupled to a 140 mm focal length spectrometer equipped a cooled CCD camera. Raman spectra of MPT sample obtained in different positions are reported in Figure 1, evidencing the simultaneous presence of Raman lines arising from anatase phase (Raman-active modes $B_{1g}$ and $E_g$) and from rutile phase (Raman-active modes $E_g$ and $A_{1g}$). These results indicate that micro-crystallites of both $TiO_2$ crystal phases were present in MPT samples over a region of bout 20 μm in size (corresponding approximately to the diameter of sampled area for each Raman spectrum).

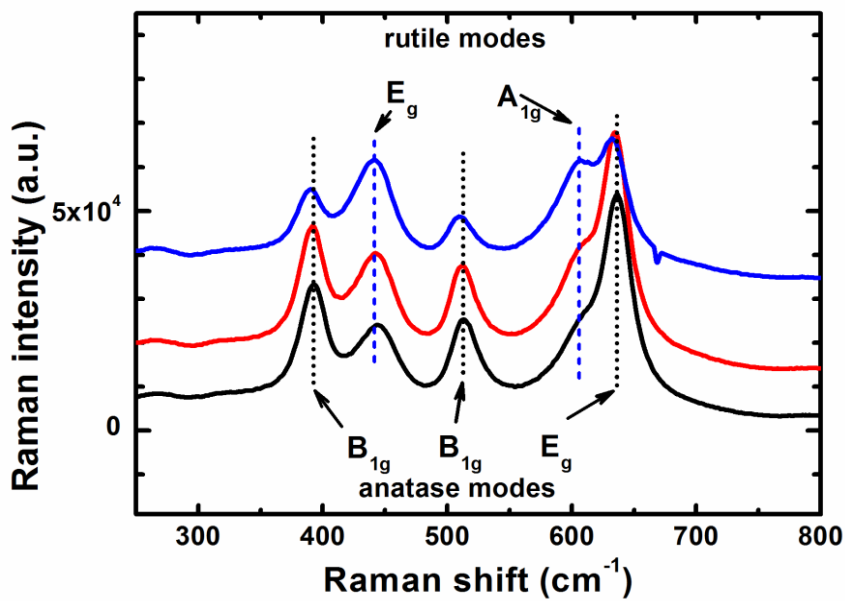

**FIG. 1.** Micro-Raman spectra obtained in different (random) surface location of an MPT sample. The intensity peaks originating from Raman-active modes by black dotted lines (anatase modes) and by blue dashed lines (rutile modes).

Room temperature PL measurements were carried out using the ultraviolet emission line (wavelength 325 nm) of a He-Cd laser as excitation source. The samples were positioned in a home-built sealed stainless steel chamber equipped with a quartz window kept in dry conditions to avoid effects on PL emission due to humidity. He-Cd laser beam impinged on the sample surface at an angle of incidence of about 45°. Time duration of samples excitation (laser exposure time) was controlled by an external shutter. The photoluminescence light diffusing from the sample surface was collected through an achromatic confocal lens system and sent to a 320 mm focal length spectrometer coupled with a thermo-cooled CCD camera via optical fiber bundle. Representative PL spectra of a MPT sample measured in dry $N_2$ (black curve) and in a 20%/80% $O_2/N_2$ mixture (red curve) are reported in Figure 2. The observation of both the VIS and NIR bands confirms the compresence of both anatase and rutile phase. Moreover, the anti-correlated behavior toward $O_2$ is also clearly evidenced. The regions highlighted in the graph indicate the wavelength interval $\Delta\lambda_V$ and $\Delta\lambda_N$ that we used to calculate $\varphi_V$ and $\varphi_N$ by numerical integration of the experimental PL intensities, *i.e.*:

$$\varphi_V = \int_{\Delta\lambda_V} \phi(\lambda) d\lambda; \quad \varphi_N = \int_{\Delta\lambda_N} \phi(\lambda) d\lambda \quad (8)$$

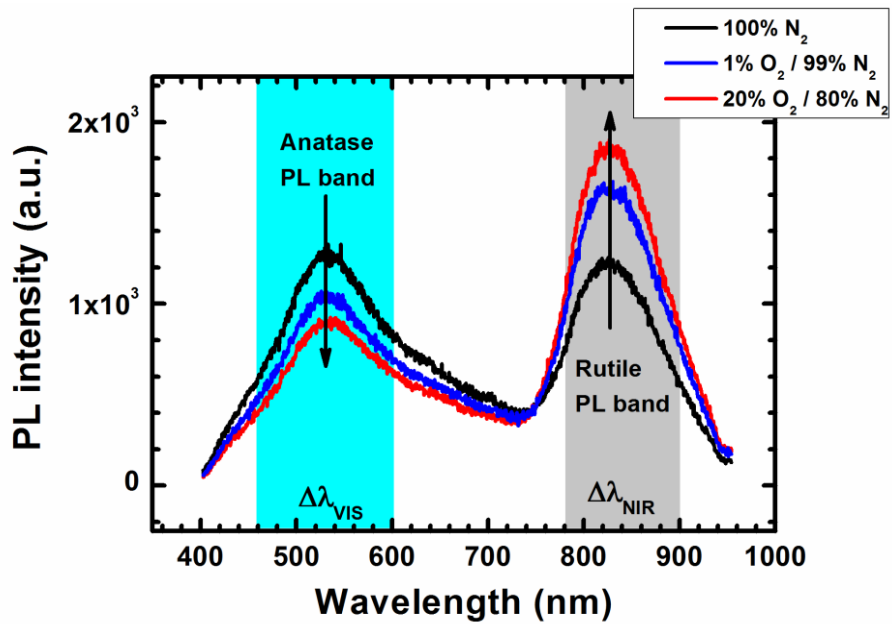

**FIG. 2.** Room temperature PL spectra of an MPT sample acquired after stabilization in dry $N_2$ (black curve) and in a mixture of dry $O_2$ and dry $N_2$ (black curve: 100% $N_2$, blue curve: 1% $O_2$ / 99% $N_2$, red curve: 20% $O_2$ / 80% $N_2$). The vertical arrows indicate the direction of PL intensity modifications induced by interaction with $O_2$, evidencing the anti-correlated behavior of the two crystalline phases. The wavelength intervals used to quantify the VIS-PL and NIR-PL intensities are highlighted.

In the above equation, $\phi(\lambda)$ is the spectral PL intensity (vs. emission wavelength $\lambda$) and the wavelength intervals $\Delta\lambda_V$ and $\Delta\lambda_N$ extend from 460 nm to 600 nm and from 780 nm to 900 nm, respectively.

PL intensity measurements on MPT and reference single-phase samples were performed during exposures to $O_2/N_2$ flows of different $O_2$ concentrations, alternating to sample recovery in $N_2$ flows. The $O_2$ concentration in the test chamber was maintained at the desired values by means of a computer-controlled mass flow control system, mixing and flowing dry $N_2$ (99.9995% purity) and $O_2$ at a constant total flow rate of 250 sccm. A bespoke home-prepared LabVIEW control program was used in order to monitor the gas flow and the laser power, control the external shutter, start the PL acquisition and record the acquired spectra in real-time. The time duration of each gas step was 30 minutes. Tested $O_2$ concentration were $[O_2]$ = 1%, 2%, 5% and 20%. PL acquisition were carried out each 30 second at an integration time of 1 second.

Figure 3a shows the PL intensity map (contour plot) vs. time for the overall experiment carried out on a representative MPT sample, evidencing at a glance the anti-correlated behavior of anatase and rutile PL intensities exposed to $O_2$. In Figure 3b we show the dynamical behavior of PL intensity $\varphi_N(t)$ and $\varphi_V(t)$ versus time (top, blue curve: NIR-PL $\varphi_N(t)$, bottom, red curve: VIS-PL $\varphi_V(t)$). The quantities $\varphi_N(t)$ and $\varphi_V(t)$ are determined according to Equation (8) and corresponding to the $\phi(\lambda)$ data reported in Figure 3a. Experimental runs using the same experimental parameters were also carried out on single-phase (rutile and anatase) samples (PL intensity results not shown here). Next, the data obtained for MPT and single-phase samples were used to calculate the instantaneous (time-dependent) PL responsivity appropriate for each sample, namely:

$$\begin{aligned} \text{MPT:} \quad & \bar{R}(t) = 1 - \left(\varphi_N^0/\varphi_V^0\right) \cdot \frac{\varphi_V(t)}{\varphi_N(t)} \\ \text{Rutile:} \quad & R_E(t) = 1 - \frac{\varphi_N^0}{\varphi_N(t)} \quad (9) \\ \text{Anatase:} \quad & R_Q(t) = 1 - \frac{\varphi_V(t)}{\varphi_V^0} \end{aligned}$$

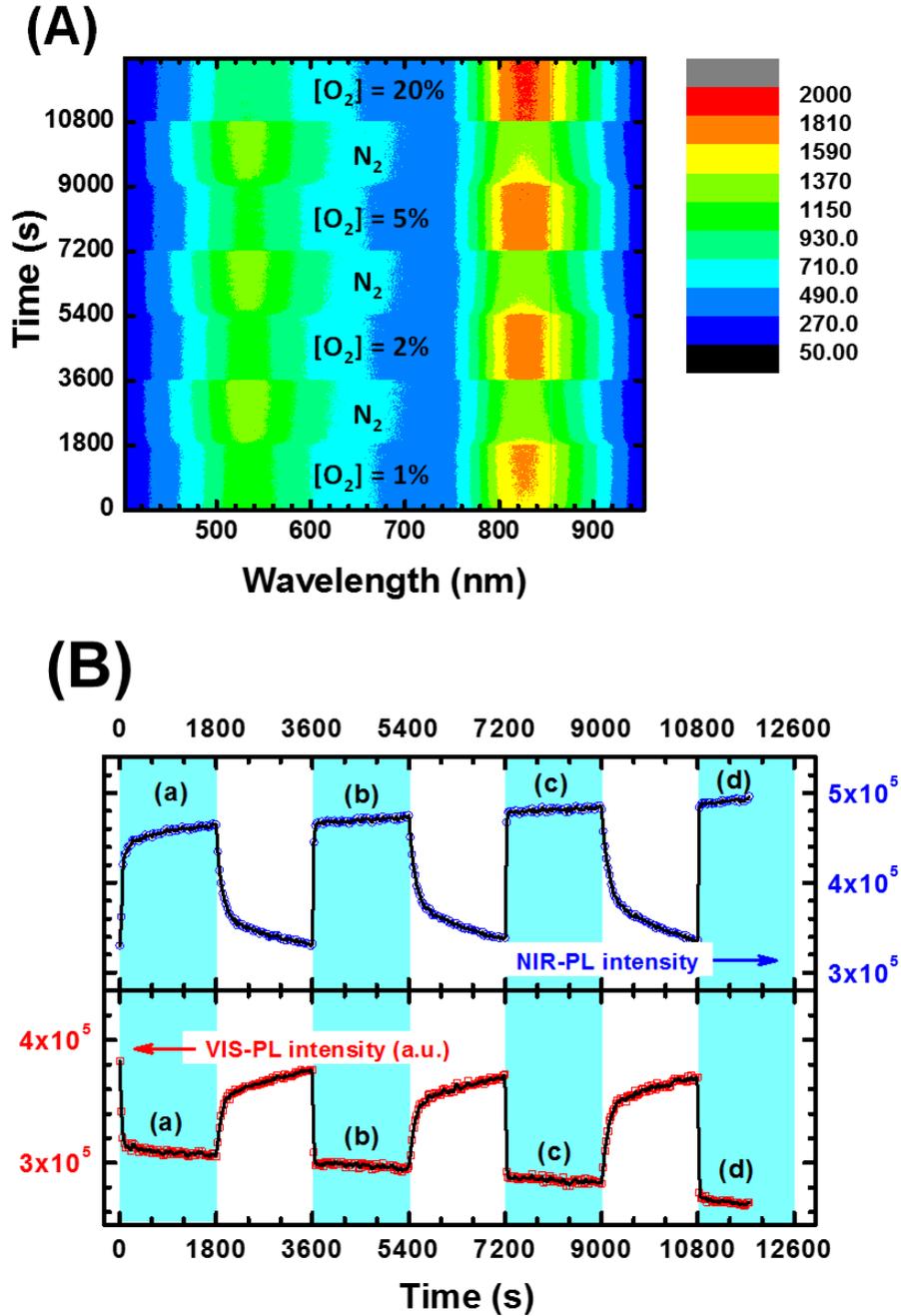

**FIG. 3.** (A): Contour plot of PL spectral intensity measured throughout the exposure to $O_2/N_2$ mixtures with different $O_2$ concentrations (indicated by [$O_2$]). (B) NIR-PL intensity $\varphi_N$ (top) and VIS-PL intensity $\varphi_V$ (bottom) extracted from the data reported in Figure 3A according to the definition in Equation (8). The $O_2/N_2$ mixtures were flown in the test chamber for 30 minutes (1800 s) starting from t = 0, alternating to 100% $N_2$ flows. The colored regions in Figure 3B indicate the presence of $O_2/N_2$ at the tested concentration of 1% (a), 2% (b), 5% (c) and 20% (d).

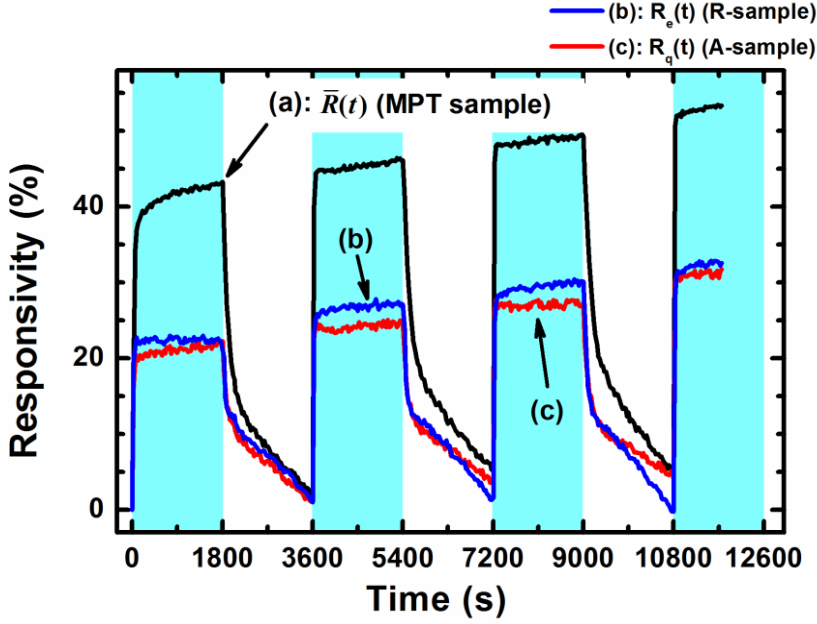

**FIG. 4.** Time dynamics of optical responsivities $\bar{R}$ (curve (a), MPT sample), $R_E$ (curve (b), rutile sample) and $R_Q$ (curve (c), anatase sample) toward different $O_2$ concentrations. For the explicit expression of the three responsivities, see Equation (9). For [$O_2$] values, see the caption of figure 3.

The results are reported in Figure 4. Responsivities ranging from 20-25% were observed for both single-phase samples (red curve: anatase; blue curve. rutile), while responsivities obtained by using the NIR/VIS PL ratio in MPT ranged from about 40% to 53%, meaning a doubling of the responsivity achieved through the ratiometric approach.

In order to obtain a sensor calibration curve and estimate the limit of detection, further measurements at lower $O_2$ concentrations have been performed. The experimental ratiometric responsivities for MPT $\bar{R}(\%)$ vs. [$O_2$] are reported in Figure 5. The horizontal dashed line represents the 3σ noise level noise level, where σ is the standard deviation of values obtained by repeated measurement of $\bar{R}$ in absence of analyte. The dashed blue curve represents a best-fit of the experimental responsivities using an empirical Toth-like[29] equation, i.e. $\bar{R} \propto b[O_2]/(1+b[O_2]^\alpha)$, where α is an exponent close to 1 (the case α = 1 corresponds to the Langmuir equation). By extrapolating the intersection point between the calibration curve and the 3σ noise level, we estimated a limit of detection of [$O_2$] = 0.009%, = 90 ppm.

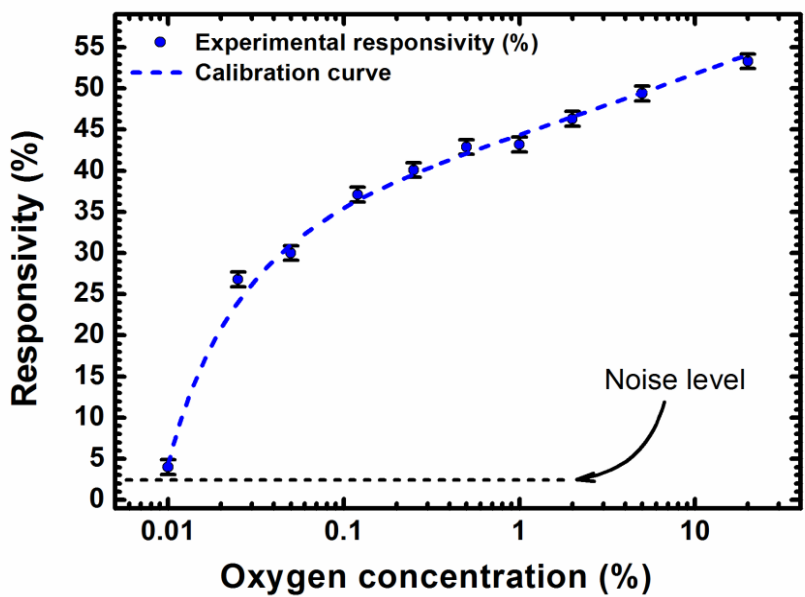

**FIG. 5**. Sensor calibration curve for the MTP pellets. The blue dots represent the experimental responsivities, the dashed black line is the noise level at three standard deviations (3σ), and the blue curve is a fit of the experimental data using a Toth-like power law. Extrapolating the intersection point between the blue curve and the 3σ noise level, a limit of detection of 90 ppm is determined.

In conclusion, we show that the peculiar and anti-correlated PL response towards $O_2$ exposure exhibited by the anatase and rutile $TiO_2$ polymorphs can be exploited using mixed-phase $TiO_2$ samples as efficient optical $O_2$ gauges exhibiting a ratiometric-enhanced optical responsivity. More precisely, the simultaneous availability of the PL responses of anatase and rutile $TiO_2$ and the detection of their ratio as transduction signal allow determining an experimental responsivity that is by construction always larger than the one obtainable in by single-phase $TiO_2$. We give here a simple proof-of-concept by testing MPT samples obtained by mixing commercial powders of different crystalline phase. Comparison of the results obtained using the NIR/VIS PL ratio with those obtained on reference single-phase samples (fabricated through the same $TiO_2$ nanopowders) indicates a two-fold enhancement of the optical responsivity provided by the ratiometric approach.

In addition to the peculiar ratiometric-enhanced optical responsivity, we underline that at least three favorable properties can be envisaged in the development of $O_2$ optical probes based on mixed-phase $TiO_2$ and on the ratiometric approach proposed here. Firstly, we mention the issue of commercial availability of the sensing material. In fact, emerging applications (e.g. food packaging) often require high quantities of the indicators at very competitive price. Unfortunately, only few $O_2$ indicators (e.g. Ru(III) polypyridyl complexes and Pt(II) and Pd(II) complexes with some porphyrins)[30] are available at acceptable prices, while instead, $TiO_2$ nanoparticles can be massively produced by low costs synthesis routes, such as flame processes[31] or sol-gel synthesis[32]. Secondly, we point out that employing a dual-luminescent probe obtained through mixing two polymorph of a same material allows exploiting the ratiometric approach *plus* the responsivity enhancement without having to tackle possible issues (e.g. different thermal stability, different range of linearity vs. excitation intensity due to unequal photobleaching) related to mixing of different components. Finally, it has to be noted that luminescent dyes used as $O_2$ optical indicators in biological studies can be toxic to the cells or can be altered by the chemical interference with the biological environment. Therefore, their dispersion in the cell medium has to be avoided and supporting biofriendly matrixes have to be developed, suited to carry and/or encapsulate them[33,34]. $TiO_2$ particles are instead self-standing, possibly avoiding the need to provide such a suitable immobilization support.

**Acknowledgements**

This work has been partially supported by the Italian Ministry for University and Research through the FIRB Project No. RBAP115AYN ("Oxides at the nanoscale: multifunctionality and applications"), the PON 2007-2013 Project "CeSMA" (Centro di Servizi di Misure Avanzate, Asse I, Obiettivo Operativo 4.1.1.4) and by the Italian Ministry of Foreign Affairs and International Cooperation (NANOGRAPH Project).

**References**

[1] J.W. Fergus, J. Mater. Sci. **38**, 4259 (2003).
[2] A.D. Brailsford, M. Yussouff, E.M. Logothetis, T. Wang, and R.E. Soltis, Sens. Actuators B Chem. **42**, 15 (1997).
[3] A.D. Brailsford, M. Yussouff, and E.M. Logothetis, Sens. Actuators B Chem. **44**, 321 (1997).
[4] C. Preininger, I. Klimant, and O.S. Wolfbeis, Anal. Chem. **66**, 1841 (1994).
[5] Y. Ma, D. Liang, J. Liu, K. Axcrona, G. Kvalheim, T. Stokke, J.M. Nesland, and Z. Suo, PLoS ONE **6**, e29170 (2011).
[6] S. Pennacchietti, P. Michieli, M. Galluzzo, M. Mazzone, S. Giordano, and P.M. Comoglio, Cancer Cell **3**, 347 (2003).
[7] A.L. Harris, Nat. Rev. Cancer **2**, 38 (2002).
[8] J.D. Crapo, Annu. Rev. Physiol. **48**, 721 (1986).
[9] L.L. Mantell and P.J. Lee, Mol. Genet. Metab. **71**, 359 (2000).
[10] R. Ramamoorthy, P.K. Dutta, and S.A. Akbar, J. Mater. Sci. **38**, 4271 (2003).
[11] Y. Amao, Microchim. Acta **143**, 1 (2003).
[12] M. Nič, J. Jirát, B. Košata, A. Jenkins, and A. McNaught, editors, in *IUPAC Compend. Chem. Terminol.*, 2.1.0 (IUPAC, Research Triangle Park, NC, 2009).

[13] J.M. Lauerhaas, G.M. Credo, J.L. Heinrich, and M.J. Sailor, J. Am. Chem. Soc. **114**, 1911 (1992).

[14] J.M. Lauerhaas and M.J. Sailor, Science **261**, 1567 (1993).

[15] C. Baratto, G. Faglia, G. Sberveglieri, Z. Gaburro, L. Pancheri, C. Oton, and L. Pavesi, Sensors **2**, 121 (2002).

[16] I.A. Levitsky, Sensors **15**, 19968 (2015).

[17] A. Bismuto, S. Lettieri, P. Maddalena, C. Baratto, E. Comini, G. Faglia, G. Sberveglieri, and L. Zanotti, J. Opt. Pure Appl. Opt. **8**, S585 (2006).

[18] S. Lettieri, M. Causà, A. Setaro, F. Trani, V. Barone, D. Ninno, and P. Maddalena, J. Chem. Phys. **129**, 244710 (2008).

[19] S. Lettieri, L. Santamaria Amato, P. Maddalena, E. Comini, C. Baratto, and S. Todros, Nanotechnology **20**, 175706 (2009).

[20] S. Lettieri, A. Setaro, C. Baratto, E. Comini, G. Faglia, G. Sberveglieri, and P. Maddalena, New J. Phys. **10**, 43013 (2008).

[21] D. Valerini, A. Cretì, A.P. Caricato, M. Lomascolo, R. Rella, and M. Martino, Sens. Actuators B Chem. **145**, 167 (2010).

[22] J.R. Sanchez-Valencia, M. Alcaire, P. Romero-Gómez, M. Macias-Montero, F.J. Aparicio, A. Borras, A.R. Gonzalez-Elipe, and A. Barranco, J. Phys. Chem. C **118**, 9852 (2014).

[23] E. Orabona, D. Pallotti, A. Fioravanti, S. Gherardi, M. Sacerdoti, M.C. Carotta, P. Maddalena, and S. Lettieri, Sens. Actuators B Chem. **210**, 706 (2015).

[24] H. Xu, J.W. Aylott, R. Kopelman, T.J. Miller, and M.A. Philbert, Anal. Chem. **73**, 4124 (2001).

[25] C. Wu, B. Bull, K. Christensen, and J. McNeill, Angew. Chem. Int. Ed. **48**, 2741 (2009).

[26] H. Hochreiner, I. Sanchezbarragan, J. Costafernandez, and A. Sanzmedel, Talanta **66**, 611 (2005).

[27] D.K. Pallotti, E. Orabona, S. Amoruso, C. Aruta, R. Bruzzese, F. Chiarella, S. Tuzi, P. Maddalena, and S. Lettieri, J. Appl. Phys. **114**, 43503 (2013).

[28] D. Pallotti, E. Orabona, S. Amoruso, P. Maddalena, and S. Lettieri, Appl. Phys. Lett. **105**, 31903 (2014).

[29] J. Tóth, Adv. Colloid Interface Sci. **55**, 1 (1995).

[30] M. Quaranta, S.M. Borisov, and I. Klimant, Bioanal. Rev. **4**, 115 (2012).

[31] H.K. Kammler, L. Mädler, and S.E. Pratsinis, Chem. Eng. Technol. **24**, 583 (2001).

[32] D.P. Macwan, P.N. Dave, and S. Chaturvedi, J. Mater. Sci. **46**, 3669 (2011).

[33] E. Monson, M. Brasuel, M. Philbert, and R. Kopelman, Biomed. Photonics Handb. **9**, (2003).

[34] Y. Cao, Y.-E.L. Koo, S.M. Koo, and R. Kopelman, Photochem. Photobiol. **81**, 1489 (2005).